\documentclass{PoS}
\usepackage{graphicx}
\title{Black holes in general relativity}
\ShortTitle{Black holes in general relativity}
\author{\speaker{Matt Visser}%
      \thanks{This research was supported by the Marsden Fund administered by the Royal Society of New Zealand.}\\
       School of Mathematics, Statistics, and Operations Research, 
Victoria University of Wellington\\
       E-mail: \email{matt.visser@msor.vuw.ac.nz}}

\abstract{What is going on (as of August 2008) at the interface between theoretical general relativity, string-inspired models, and observational astrophysics? Quite a lot. In this mini-survey I will make a personal choice and focus on four specific questions: Do black holes ``exist''? (For selected values of the word ``exist''.) Is black hole formation and evaporation unitary? Can one mimic a black hole to arbitrary accuracy? Can one detect the presence of a horizon using local physics?

\bigskip
arXiv:0901.4365 [gr-qc]; 1 February 2009; 5 February 2009; \LaTeX-ed \today.
 }
\FullConference{Black Holes in General Relativity and String Theory\\
		 24--30 August, 2008\\
		 Veli Losinj, Croatia}
\begin{document}
\section{Introduction}
\noindent
This workshop deals with a large number of topics:
\begin{itemize}
\item 
Astrophysical Black Holes  (from super-massive to stellar);
\item
Primordial and Mini Black Holes;
\item
Black Hole Entropy;
\item
Information Paradox;
\item
Asymptotic Symmetries;
\item
Anomalies;
\item
Attractor Mechanism;
\item
Holography;
\item
ADS/CFT correspondence.
\end{itemize}
Instead of trying (uselessly) to address each of these topics in turn, I will use this mini-survey to address four specific questions, these questions representing my personal choice regarding the currently important questions of black hole physics:
\begin{itemize}
\item Do black holes ``exist''?
\item Is black hole formation and evaporation unitary? 
\item Can one mimic a black hole to arbitrary accuracy? 
\item Can one detect the presence of a horizon using local physics?
\end{itemize}

\section{Do black holes ``exist'' ?}

\noindent
This innocent question is more subtle than one might expect, and the answer depends very much on whether one is thinking as an observational astronomer, a classical general relativist, or a theoretical physicist.

\paragraph{Observational astronomer:}

Astronomers have certainly seen things that are small, dark, and heavy. But are these small, dark, heavy objects  really black holes in the sense of classical general relativity? Accretion disks, and in particular their inner cutoff radius,  probe the spacetime geometry surrounding black hole candidates down to the innermost stable circular orbit [ISCO] $2m/r \approx 1/3$. (See, for example,~\cite{Kerr,Fabian}.) Furthermore, advection dominated accretion flows [ADAFs] are commonly interpreted as probing the spacetime geometry down to $2m/r \lesssim 1$, though there are still some disagreements on this point within the community. (See, for example,~\cite{accretion, ADAF, ADAF2, ADAF3, ADAF4}.) Certainly, everything observed so far is compatible with the standard Schwarzschild and/or Kerr spacetimes. (See, for example,~\cite{Fabian, Melia, van-Putten}.)

\paragraph{Classical general relativist:}

Eternal black holes certainly exist mathematically, as stationary vacuum solutions to the Einstein equations. (See, for example,~\cite{kerr-intro, kerr-on-kerr, carter-on-kerr, robinson-on-kerr}, or any of the many standard textbooks in general relativity~\cite{textbooks}.)
Furthermore classical astrophysical black holes (future event horizons) certainly exist mathematically as the end result of classical collapse based on certain physically plausible equations of state.  (See, for example,~\cite{Harrison-et-al}.)

\paragraph{Theoretical physicist:}

But one can always argue that we have not seen direct observational evidence of the event horizon~\cite{Marek}, and unless and until we do so one retains some freedom to postulate new and different near-horizon physics. In fact doing so is not entirely perverse, in that a well-thought-out alternative to standard near-horizon physics can give observational astronomers specific ideas of what to look for. 

\bigskip

More radically one could just pick one's favourite ``problem'' or ``oddity'' related to black hole physics, and use it as an excuse to speculate. One of the most abused issues in this regard (which still has a nugget of physics hiding at its core) is the relationship between unitarity, information, entropy, and black hole physics.

\section{Is black hole formation and evaporation unitary?}

\begin{figure}[!htbp]
\begin{center}
\includegraphics[width=8cm]{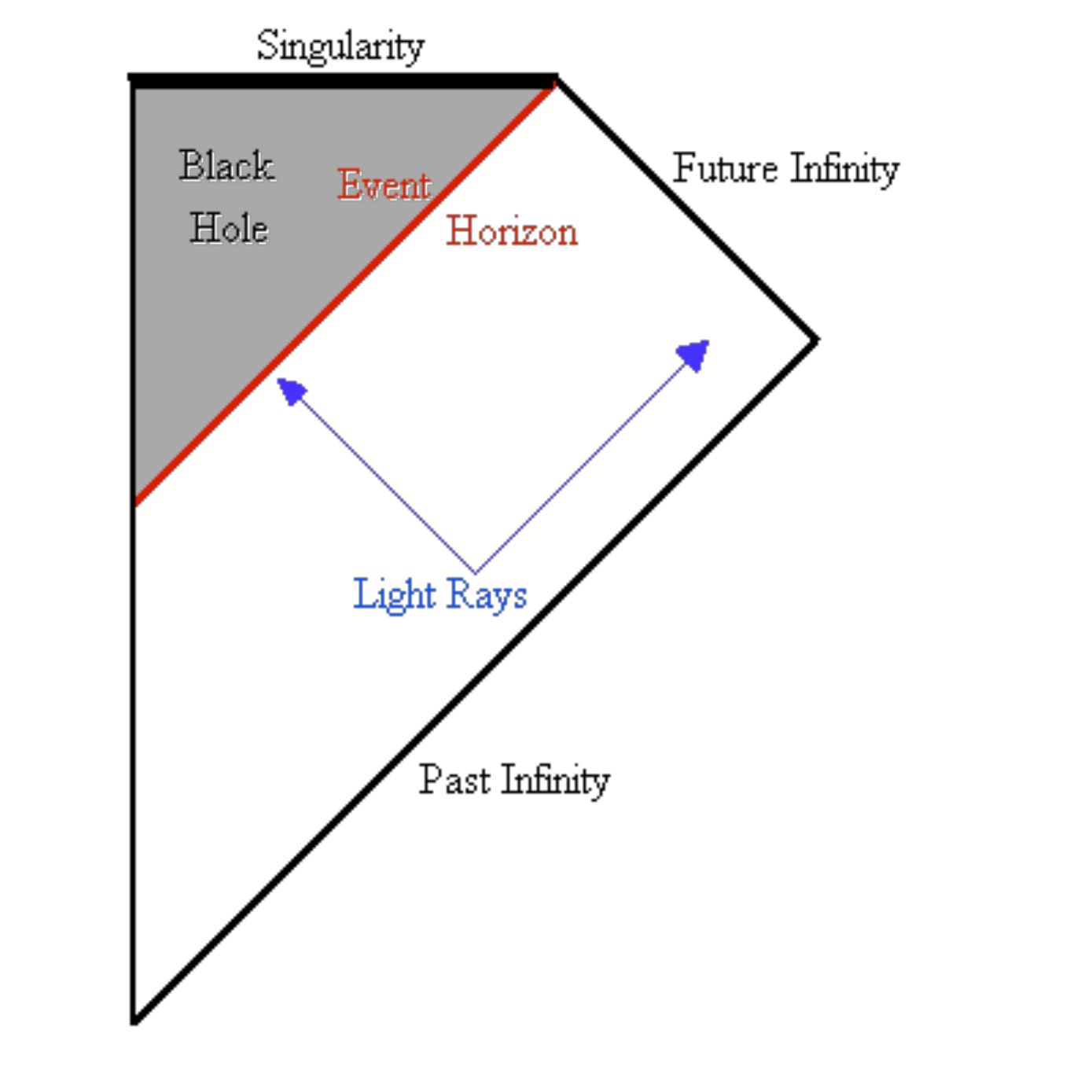}
\caption{Strict event horizon (absolute horizon).}
\label{F:strict}
\end{center}
\end{figure}

The most important thing to realise is that information ``loss'' is (in and of itself) not a problem, it is merely a ``feature''. Information loss is just one of those things you have to live with if you accept the standard Carter--Penrose diagram for the causal structure of stellar collapse, see figure~\ref{F:strict}. For all practical purposes: Information loss $\Leftrightarrow$ non-unitary evolution in the domain of outer communication  $\Leftrightarrow$ existence of an event horizon.

As soon as one has an event horizon, one has an inaccessible region, and an external observer must then ``trace over'' the states hidden in that inaccessible region, and for that observer pure states become density matrices. Thus an event horizon (absolute horizon) automatically leads to non-unitary evolution --- at least as seen from the outside. 

But is ``event horizon'' the right concept to be using? There are many other possible definitions of horizon: apparent~\cite{textbooks, apparent, Lorentzian},  dynamical~\cite{dynamical}, and/or trapping~\cite{trapping} horizons, that may make more physical sense. Even classically,  event horizons are seriously diseased in numerical general relativity, simply because they are so difficult to find with any certainty --- local or quasi-local definitions of horizon are often preferable from a purely pragmatic point of view~\cite{numerics1, numerics2}.

\begin{figure}[!htbp]
\begin{center}
\includegraphics[width=8cm]{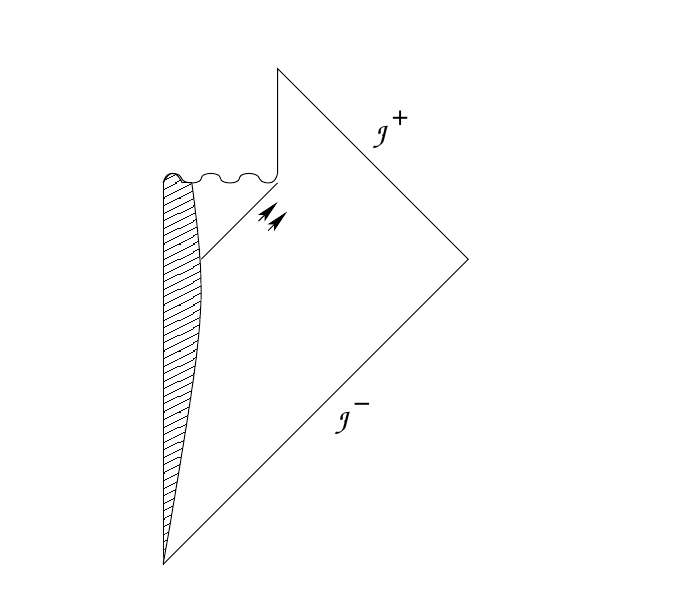}
\caption{Standard Carter--Penrose diagram for an evaporating black hole.}
\label{F:standard}
\end{center}
\end{figure}

Furthermore, once you add semiclassical quantum physics, specifically Hawking evaporation, the issues again become more complicated, leading to the so called ``information paradox'' --- what (quantitatively) happens to the information you thought you had lost once the evolving black hole reaches its final state? (Whatever that final state happens to be.)   If you believe that Hawking evaporation is unitary (as seen from our own asymptotically flat domain of outer communication), then there had better not be any any event horizon in the spacetime. To quote Stephen Hawking in the abstract to his talk delivered at GR17:
\begin{quotation}
``The way the information gets out seems to be that a true event horizon never forms,  just an apparent horizon.''
\end{quotation}
(The actual talk given at GR17 has never been published, even informally, though transcripts can be found on the web. The scientific publication closest in content to that talk is~\cite{Stephen}.) Largely independent of one's favourite model for the ``quantum theory that would be gravity'', whether loop-based, string-based, or something else, the last few years has seen a consensus shift to where many if not most physicists now seem to think that Hawking evaporation is unitary.
Thus the  event/ absolute/ apparent/ trapping/ dynamical horizon distinction may be more than just a matter of mathematical convenience --- it may be critically important to the underlying physics of black hole formation and evaporation. (For earlier somewhat related ideas, see, for instance,~\cite{Hajicek3}.)

Consider the standard Carter--Penrose diagram for an evaporating black hole, figure~\ref{F:standard}. In this standard picture there is always a region of ``don't ask because we certainly cannot tell'' associated with the (vertical) timelike line emerging from the endpoint of the evaporation process.  This is the source of the famous tri-chotomy: Is the endpoint of Hawking evaporation
\begin{itemize}
\item complete evaporation?
\item a timelike (naked) singularity?
\item a stable remnant?
\end{itemize}
As long as one insists on the black hole being defined by an event horizon (absolute horizon), then the standard diagram of figure~\ref{F:standard} is unavoidable --- as is the associated information loss and non-unitary evolution.

\begin{figure}[!htbp]
\begin{center}
\includegraphics[width=8cm]{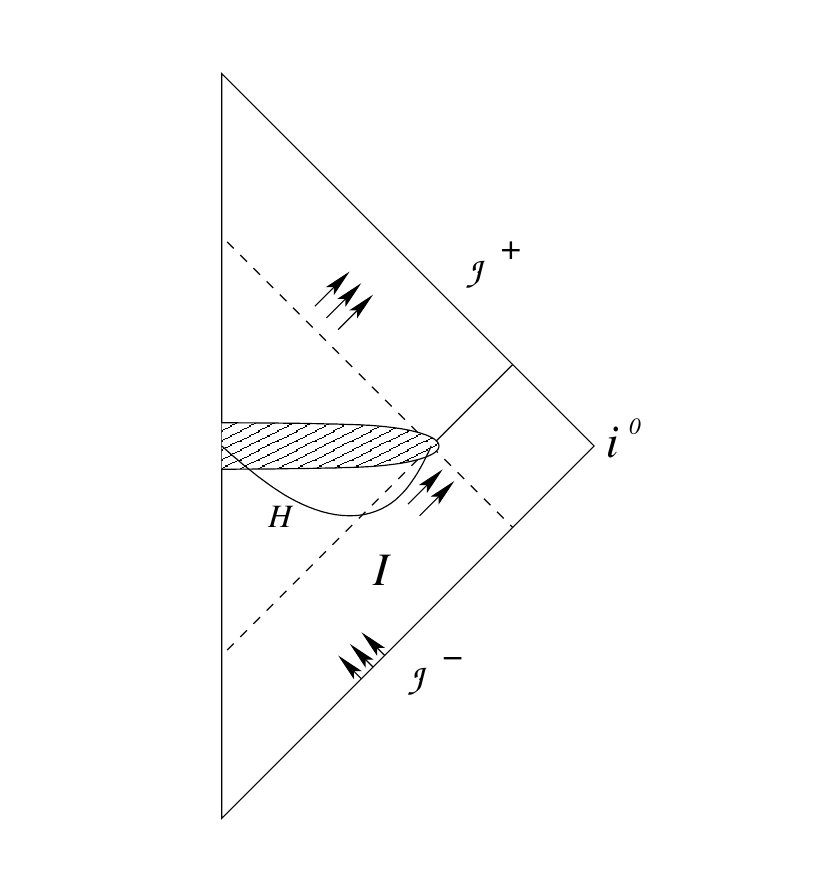}
\caption{Ashtekar--Bojowald version of the Carter--Penrose diagram for an evaporating black hole.}
\label{F:AB}
\end{center}
\end{figure}

As an counterpoint to the standard Carter--Penrose diagram,  Ashtekar and Bojowald~\cite{Ashtekar-Bojowald}  have suggested replacing the spacelike singularity by a region of Planckian curvature --- leading to an alternative Carter--Penrose diagram for the causal structure of black hole evaporation.  While Ashtekar and Bojowald were working within the framework of loop quantum gravity when they drew their version of the Carter--Penrose diagram, figure~\ref{F:AB}, it is critically important to realize that qualitatively the same sort of causal diagram might be compatible with a unitarily evolving string-based model of black hole evaporation. Depending on one's favourite model for ``quantum gravity'' the shaded region of Planckian curvature might be viewed as a ``loop network'', ``strongly interacting string muck'', or more prosaically as a region of violently fluctuating spacetime.

While there is no event horizon in the Ashtekar--Bojowald diagram, there typically is an apparent or dynamical horizon. There definitely is a ``Planck horizon'' or ``reliability horizon''. Once one crosses the lower-left dashed line, (which replaces the notion of event horizon), one is guaranteed a personal encounter with Planck scale physics. While this is likely to be just as fatal as running into a spacelike singularity, the highly curved (highly fluctuating, highly discrete) shaded region is now no longer an absolute barrier to unitary evolution. (Related ideas have also arisen in the study of chronology horizons and the ``chronology protection conjecture''~\cite{reliability, reliability2, cpc1, cpc2}. For some older ideas on singularity avoidance, see, for instance~\cite{Hajicek2}.)

\begin{figure}[htbp]
\begin{center}
\includegraphics[width=8cm]{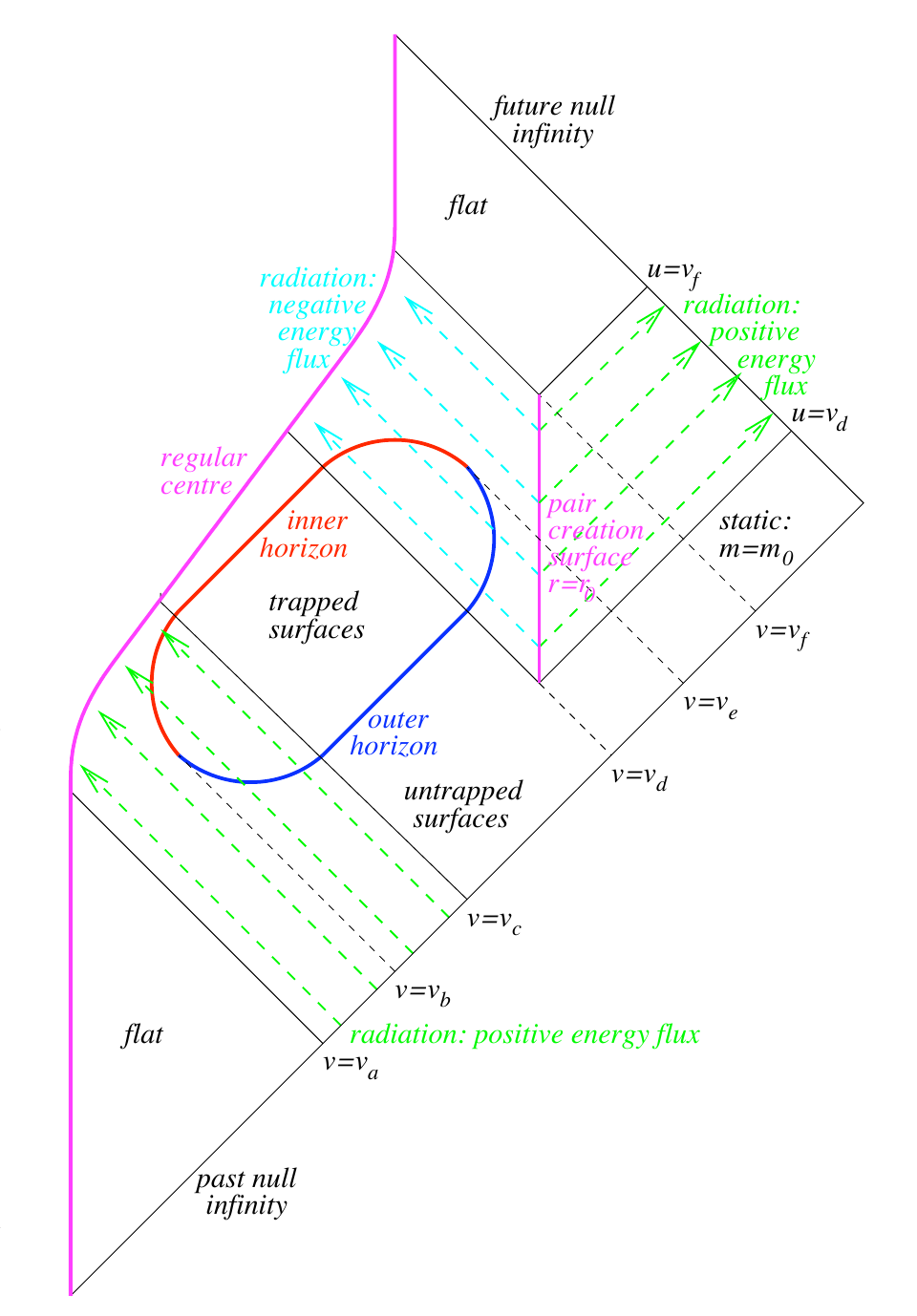}
\caption{Hayward version of the Carter--Penrose diagram for an evaporating black hole.}
\label{F:SH}
\end{center}
\end{figure}

Yet another causal diagram for an evaporating black hole is that due to Hayward~\cite{Hayward}. At a technical level the main difference is that Hayward prefers to work with trapping horizons~\cite{trapping}, while Ashtekar and Bojowald prefer to work with dynamical horizons~\cite{dynamical}. While the diagram in figure~\ref{F:SH} at first glance looks rather different from that in figure~\ref{F:AB}, remember that all these causal diagrams are conformal diagrams --- they depend to some extent on one's choice of coordinates, and additionally are specified only up to a conformal factor that can be used to expand or contract parts of the diagram depending on what aspect of the physics one is most interested in. In particular, the purple line in figure~\ref{F:SH} labelled ``regular centre'' is timelike, and with a suitable choice of coordinates can be ``straightened out'' so that it looks like the vertical line in figure~\ref{F:AB}.  Furthermore, in figure~\ref{F:SH} no attempt has been made to identify the region of Planck-scale curvature. Despite appearances, the two causal diagrams of figures \ref{F:AB} and \ref{F:SH} are to a large extent compatible with each other, and both are compatible with the unitary evolution and evaporation of black holes.
This qualitative result does not depend on whether you choose to believe in loops or strings, or indeed any other candidate theory of ``quantum gravity'' that has Einstein gravity as an approximate low-energy limit.

\begin{figure}[!htbp]
\begin{center}
\includegraphics[width=7cm]{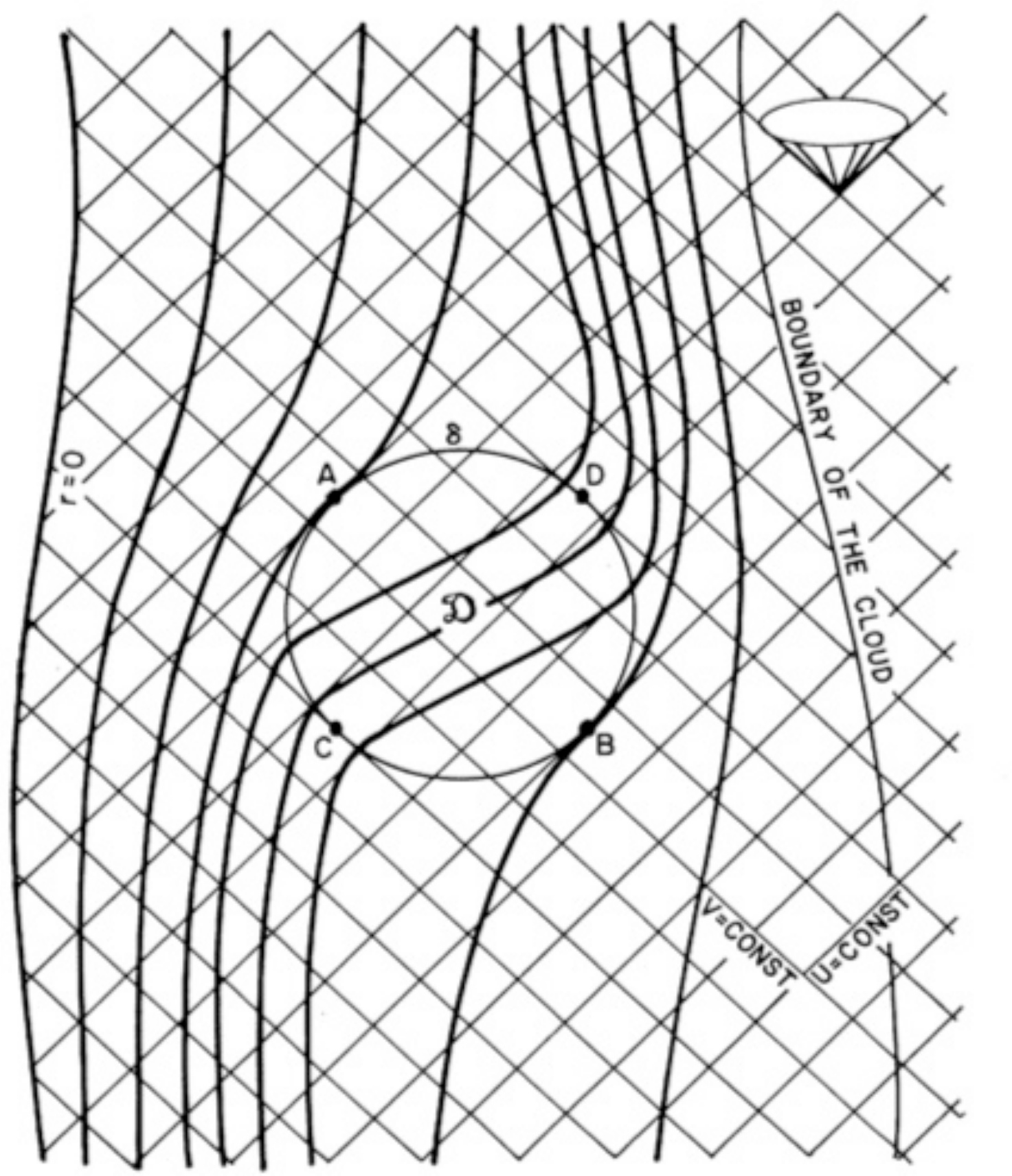}
\caption{Roman--Bergmann double-null causal diagram for a regular collapsing star.}
\label{F:BR1}
\end{center}
\end{figure}

There is also some very nice considerably older work by Roman and Bergmann~\cite{Roman}, where they considered the possibility of obtaining apparent horizons without an event horizon.  While they were not specifically interested in Hawking radiation, being more interested in ideas of stellar collapse with a regular centre, many of the qualitative features of that article now resonate in the Hayward and Ashtekar--Bojowald causal diagrams. In particular, the curve labelled  $r=0$ on the left of  figure~\ref{F:BR1} is timelike, and with a suitable choice of coordinates can be ``straightened out'' so that it looks like the vertical line in figure~\ref{F:AB}.  Furthermore, although  Roman and Bergmann adopted double null coordinates to clarify the causal structure, they did not perform the singular conformal transformation that is needed to pull asymptotic infinity in to a finite ``distance'' and so produce a Carter--Penrose diagram. Not performing this singular conformal transformation is sometimes an advantage; it can make the diagram a little more realistic when it comes to studying the metric features, (as opposed to causal features).

\begin{figure}[!htbp]
\begin{center}
\includegraphics[width=8cm]{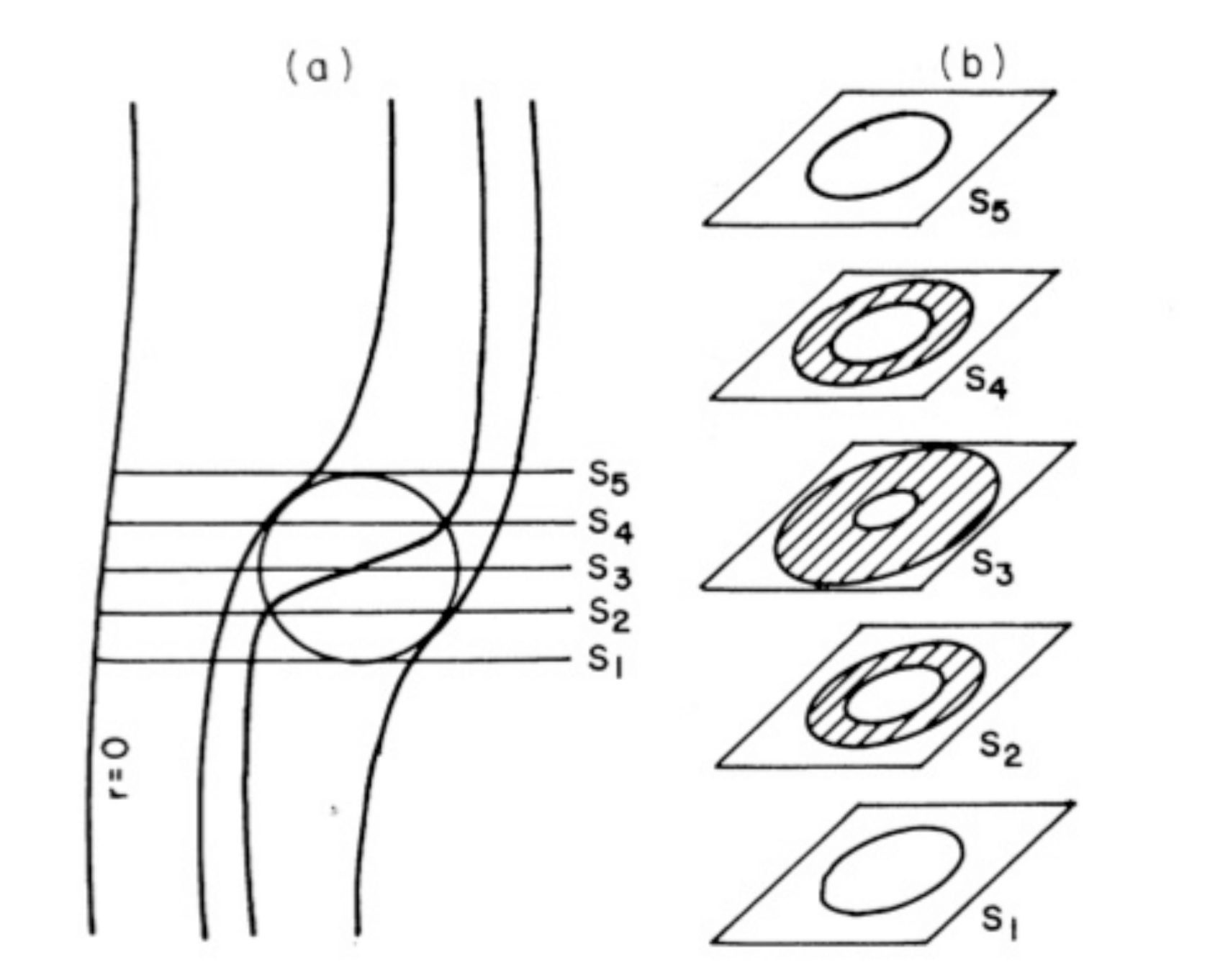}
\caption{Null geodesic expansion/contraction for a Roman--Bergmann regular collapsing star.}
\label{F:BR2}
\end{center}
\end{figure}

These ideas have now mutated and spread throughout the theoretical relativity community, and variants of these ideas can be seen, for instance, in recent work by Nielsen~\cite{Nielsen}.  I should caution the reader that just sketching a suitable causal diagram is only the first step, since by definition Carter--Penrose diagrams only capture the conformal structure of the spacetime. One should in addition think very carefully about the metrical structure --- and one should fix the conformal factor so that the spacetime \emph{geometry} is close to Schwarzschild and/or Kerr over a suitably large portion of the manifold. Furthermore, one should really then patch the resulting spacetime geometry onto a self-consistent quantum calculation leading to a Hawking-like flux near spatial infinity. Only then would one have a really compelling picture of black hole collapse and evaporation.

A key aspect of all these discussions is this critical point: \emph{You do not need an event horizon to get Hawking radiation.}  A long-lived apparent/ dynamical/ trapping horizon is more than sufficient. This is (or should be) a well-known result, dating back (at least) to Hajicek~\cite{Hajicek}, and many other workers of that period. This is one of those elementary and utterly important results that has been repeatedly forgotten, and repeatedly rediscovered, over the last 30 years~\cite{essential}.

More recently it has been realised that \emph{you do not even need apparent/ dynamical/ trapping horizons to get a Hawking-like flux} --- asymptotic approach to trapping horizon formation is quite sufficient~\cite{no-trapping1, no-trapping2}.
This naturally leads one to ask: Do we actually need ``black holes'' to do ``black hole physics''?

\section{Black hole mimics?}
\noindent
Even if we do not actually need ``black holes'' to do ``black hole physics'', can we at least be sure that we have something very similar to a black hole, a ``black hole mimic''?
\begin{itemize}
\item 
Can one avoid black hole formation with a suitably weird equation of state?
\item
Can one avoid black hole formation with semi-classical quantum effects? 
\item
Can one avoid black hole formation with ``quantum gravity''? 
\end{itemize}
The possibilities are rather tightly constrained. (There is of course the utter gibbering crackpot fringe, but names will be suppressed to protect the guilty.)  The ``physically reasonable'' alternatives to black hole formation can be  counted on the fingers of one hand. (For selected values of  ``physically reasonable''.)
This limited set of alternatives includes:
\begin{itemize}
\item 
Quark stars~\cite{quark},  Q-balls~\cite{q-ball},   strange stars~\cite{strange}.
\item
Boson-stars~\cite{boson}.
\item
Gravastars:    Mazur--Mottola variant~\cite{Mazur-Mottola},  and  Laughlin \emph{et al.}~variant~\cite{Laughlin}.
\item
Fuzz-balls:    Mathur \emph{et al.}~variant~\cite{Mathur}, and  Amati variant~\cite{Amati}.
\item
Dark stars/ Quasi-black holes~\cite{fate}, and other proposals similar in spirit~\cite{Boulware, Vachaspati}.
\end{itemize}

\paragraph{Quark stars,  Q-balls,  strange stars:}

The idea here is to change the equation of state of nuclear matter under extreme conditions~\cite{quark, q-ball, strange}:  
\begin{equation}
\hbox{
Ordinary star $\rightarrow$ white dwarf  $\rightarrow$ neutron star  $\rightarrow$ (something unexpected?) 
}
\end{equation}
The justification for the hypothesised equation of state is somewhat questionable. Note that one still has the Buchdahl--Bondi bound:  $2m/r \leq 8/9$ for any isotropic pressure profile.
So you cannot get  ``close''  to $2m/r \lesssim1$, which is a signal that you are ``close'' to forming an apparent/ dynamical/ trapping horizon, unless you are at the very least willing to adopt anisotropic stresses~\cite{anisotropic}.

\paragraph{Boson stars:} The idea here is to have a (classical or possibly semiclassical) scalar field configuration corresponding to a heavy compact object~\cite{boson}. Anisotropic pressures then ariase from the fact that the scalar Lagrangian, combined with spherical symmetry, automatically leads to an anisotropic classical stress-energy tensor. (And in view of the Buchdahl--Bondi bound,  you will certainly need the anisotropies if you want to get  ``close''  to $2m/r \lesssim1$.)

\paragraph{Gravastars:}
These are hypothetical objects where the core is de Sitter like, and the exterior is Schwarzschild like, and something odd has to happen in the region $2m/r \lesssim 1$ where the horizon would otherwise have formed~\cite{Mazur-Mottola, Laughlin}. See also~\cite{Wiltshire-and-others, Lobo, Chirenti}. In the region $2m/r \lesssim 1$ various authors have argued for:
\begin{itemize}
\item 
Guaranteed anisotropies~\cite{anisotropic};
\item
Breakdown of the spacetime manifold~\cite{Laughlin}\,?
\item
Quantum effects from the one-loop action~\cite{Mottola}\,?
\end{itemize}

\paragraph{Fuzz balls:}

Explicit calculations appear to be limited to the extremal/ near-extremal regime. The black hole  ``interior'' is replaced by ``strongly interacting string muck''. The black hole ``interior'' is not  \emph{a}  spacetime,  rather it is taken to be a \emph{superposition} of  ``spacetimes''. (And none of the individual  ``spacetimes''  in the superposition has a horizon~\cite{Mathur, Amati, Amati-et-al}.  See~\cite{Skenderis} for a survey.) 

\begin{figure}[htbp]
\begin{center}
\bigskip
\includegraphics[width=9.5cm]{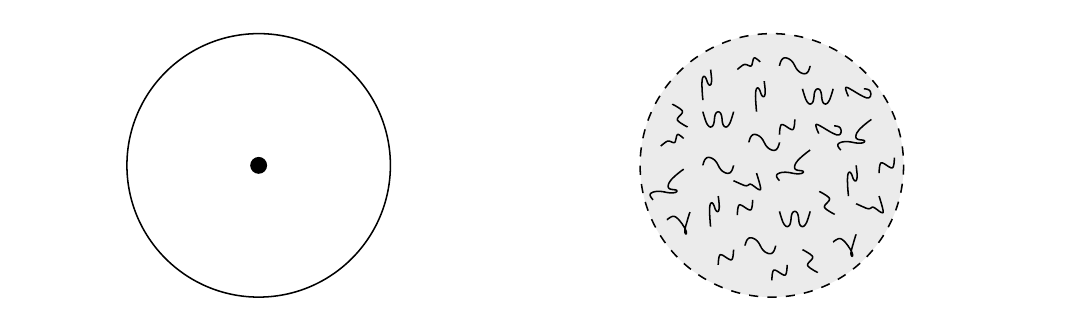}
\caption{Fuzzball picture.}
\label{F:fuzz}
\end{center}
\end{figure}

\paragraph{
Dark stars/ Quasi-black holes:}

Here the idea is to have some nonsingular configuration that mimics a black hole~\cite{fate}, see also~\cite{Boulware, Vachaspati}. Typically the configuration is continually and asymptotically collapsing, but never quite developing a horizon.  Typically the authors appeal to quantum effects as an intrinsic part of the picture. (See the accompanying article~\cite{PoS2}, and reference~\cite{fate}, for more details on the specific proposal by the current author and his collaborators.)

\newpage
\noindent
For all of these black hole mimics, there are substantial issues to address:
\begin{itemize}
\item 
Calculations are often somewhat less explicit than one would like, simply because computations are technically difficult.
\item 
There are a number of observational issues to address. For instance, rotating black hole mimics typically exhibit an ergoregion instability~\cite{ergoregion}.  There are also constraints coming from accretion observations~\cite{narayan}. 

\end{itemize}
In short, one does not have complete freedom to speculate, any model for a black hole mimic along any one of the lines described above should at some stage be carefully confronted with the observational constraints.

\section{Detecting horizons?}

\noindent
A common statement that one often encounters in the literature is this:
\begin{itemize}
\item ``Horizons are not detectable with local physics''.
\end{itemize}
The above statement is, of course, \,\emph{false}.
Note however, that it is  \emph{almost}\/  true. Two closely related, but  \emph{true}, statements are:  
\begin{itemize}
\item 
``Event horizons are sometimes not detectable with local physics'';
\item
``Apparent/ dynamical/ trapping  horizons are not  detectable with ultra-local physics''.
\end{itemize}
Note the very careful and precise phrasing of these two statements.

\bigskip

Regarding the first point, it is an elementary exercise to verify that event horizons can form in locally flat portions of Minkowski spacetime. Just let a dust shell collapse (classical dust, zero pressure). Apply Birkhoff's theorem. The region outside the shell is a portion of the Schwarzschild spacetime. The region inside the shell is a portion of Minkowski spacetime. The event horizon (and I really do mean event horizon, this statement is not true for apparent/ dynamical/ trapping  horizons) first forms at $r=0$ in flat Minkowski space where the Riemann tensor is zero, and then sweeps out to encounter the infalling dust shell just as it crosses $r=2m$. (If one insists on an explicit calculation, then it is advisable to use nonsingular coordinates, such as Painlev\'e--Gullstrand coordinates.)  The key point is this: 
\[
\hbox{Event horizon $\neq$ strong (local) gravity.}
\]
Event horizons are statements about the global geometry, even if the global geometry is sufficiently distorted to prevent light escaping to infinity, this does not tell you anything about the local strength of the gravitational field at the event horizon itself. In particular, this class of event horizons, because they occur in flat Minkowski space, are certainly not detectable (even in principle) by any local physics. (If one has additional ``metadata'' this situation can change. For instance, if one knows \emph{a priori} that the spacetime is static, then event horizons are also Killing horizons and are also apparent/ dynamical/ trapping horizons  --- as we shall soon see, this additional structure may then make them detectable using local (though not ultra-local) physics.)

Regarding the second point, spherically symmetric apparent/ dynamical/ trapping horizons in spherically symmetric spacetimes are typically associated with $2m(r)/r \approx 1$. Furthermore, at least in spherical symmetry, the quantity $2m(r)/r$  is measurable using \emph{local} (though not \emph{ultra-local}) physics. To see this, recall that in any finite-size laboratory you can measure the Riemann tensor. (Equivalently, physics with finite-range interactions is sensitive to the Riemann tensor.) But in spherical symmetry the orthonormal components 
of the Riemann tensor are linear combinations of density, radial and transverse pressures, and the quantity $2m(r)/r^3$ (which is related to the average density inside radius $r$). In fact, it is an easy exercise to check~\cite{Lorentzian} (see especially pages 27--28, and page 110, exercise 1):
\begin{eqnarray}
R^{\hat t}{}_{\hat r \hat t \hat r}  &=&  {2m(r)\over r^3} - 4\pi(\rho-p_r+2p_t);
\\
R^{\hat t}{}_{\hat \theta \hat t \hat \theta}  =  R^{\hat t}{}_{\hat \phi \hat t \hat \phi} &=&   
- {m(r)\over r^3} - 4\pi p_r;
\\
R^{\hat r}{}_{\hat \theta \hat r \hat \theta}  =  R^{\hat r}{}_{\hat \phi \hat r \hat \phi} &=&   
- {m(r)\over r^3} + 4\pi \rho;
\\
R^{\hat \theta}{}_{\hat \phi \hat \theta \hat \phi} &=&   
 {2m(r)\over r^3}.
\end{eqnarray}
Here $p_r$ is the radial pressure, $p_t$ the transverse pressure, and $\rho$ the density. 
Now the stress-energy tensor is certainly measurable using local physics, therefore  $2m(r)/r^3$  is measurable using local 
(though not ultra-local) physics. Finally, $r$ itself is just the radius of curvature of the surface of spherical symmetry passing through the point of interest, and is again measurable using local  (though not ultra-local) physics.

Combining these observations:  In spherically symmetric spacetimes $2m(r)/r$  is measurable using local (though not ultra-local) physics, and so the presence of spherically symmetric apparent/ dynamical/ trapping horizons \emph{are} detectable using local (though not ultra-local) physics.
At the risk of initiating tribal warfare:
\begin{itemize}
\item 
\emph{In spherically symmetric spacetimes, restricting attention to spherically symmetric horizons: The most physically interesting horizons, the  apparent/ dynamical/ trapping horizons, are detectable using local (though not ultra-local) physics.}
\end{itemize}
When one moves beyond spherical symmetry, either of the spacetime itself or of the trapped surfaces and/or world tubes used to define  apparent/ dynamical/ trapping horizons, then the situation becomes considerably more complex~\cite{senovilla}.

\section{Discussion}
The message that I hope readers take from this micro-survey is that there are still many subtle and interesting things going on in black hole physics. Many deep issues of principle remain, despite at least 50 years, and by some measure 90 years, work on the subject. Note also that in many cases it is worthwhile to carefully re-analyze and re-assess work from several decades ago. Sometimes, in my more cynical moments, I feel that each step forward involves 99\% of a step backwards:
\begin{quote}
``Take the red pill to remain in denial. The blue pill to accept the bleak reality...''
\end{quote}

\section*{Acknowledments}
I wish to thank Carlos Barcel\'o, Veronika Hubeny, Stefano Liberati, Jos\'e Senovilla, and Sebastiano Sonego for their comments and feedback.


\end{document}